\documentclass[aps,prl,floatfix,twocolumn,superscriptaddress,showpacs]{revtex4}
\usepackage{bm}
\usepackage{epsf}
\usepackage{amssymb}
\usepackage{amsmath}
\usepackage{graphicx}
\usepackage{rotating}
\usepackage{epsfig}
\usepackage{psfrag}
\usepackage{amsmath}
\usepackage{hyperref}
\usepackage{subfigure}

\hypersetup{
    bookmarks=true,         
    unicode=false,          
    pdftoolbar=true,        
    pdfmenubar=true,        
    pdffitwindow=true,      
    pdftitle={My title},    
    pdfauthor={Author},     
    pdfsubject={Subject},   
    pdfcreator={Creator},   
    pdfproducer={Producer}, 
    pdfkeywords={keywords}, 
    pdfnewwindow=true,      
    colorlinks=true,       
    linkcolor=red,          
    citecolor=green,        
    filecolor=magenta,      
    urlcolor=cyan           
}

\DeclareMathAlphabet{\bi}{OML}{cmm}{b}{it}

\begin{document}
\def\ea{\textit{et al.}}
\def\bF{{\mathbf F}}
\def\ba{{\mathbf a}}
\def\bB{{\mathbf B}}
\def\bE{{\mathbf E}}
\def\bj{\bm{j}}
\def\bJ{{\mathbf J}}
\def\bA{{\mathbf A}}
\def \xy{$x$--$y$ }
\def\bP{{\bf P}}
\def\bK{{\bf K}}
\def\bk{{\bf k}}
\def\bkn{{\bf k}_{0}}
\def\bx{{\bf x}}
\def\bz{{\bf z}}
\def\bR{{\bf R}}
\def\br{{\bf r}}
\def\bq{{\bf q}}
\def\bp{{\bf p}}
\def\bQ{{\bf Q}}
\def\bs{{\bf s}}
\def\bG{{\mathbf G}}
\def\bv{{\bf v}}
\def\b0{{\bf 0}}
\def\la{\langle}
\def\ra{\rangle}
\def\Im{\mathrm {Im}\;}
\def\Re{\mathrm {Re}\;}
\def\beq{\begin{equation}}
\def\eeq{\end{equation}}
\def\bea{\begin{eqnarray}}
\def\eea{\end{eqnarray}}
\def\bdm{\begin{displaymath}}
\def\edm{\end{displaymath}}
\def\bnab{{\bm \nabla}}
\def\Tr{{\mathrm{Tr}}}
\def\sfrac{\textstyle\frac}
\def\Sr{\mathrm{Sr}_2\mathrm{RuO}_4}

\title{Intrinsic Hall Effect in a Multiband Chiral Superconductor in the Absence of an External Magnetic Field}
\author{Edward~Taylor}
\affiliation{Department of Physics and Astronomy, McMaster University, Hamilton, Ontario, L8S 4M1, Canada}
\author{Catherine~Kallin}
\affiliation{Department of Physics and Astronomy, McMaster University, Hamilton, Ontario, L8S 4M1, Canada}

\date{April 10, 2012}

\begin{abstract}
We identify an intrinsic Hall effect in multiband chiral superconductors in the absence of a magnetic field (i.e., an \emph{anomalous} Hall effect).  This effect arises from interband transitions involving time-reversal symmetry-breaking chiral Cooper pairs.  We discuss the implications of this effect for the putative chiral $p$-wave superconductor, $\Sr$, and show that it can contribute significantly to Kerr rotation experiments.  Since the magnitude of the effect depends on the structure of the order parameter across the bands, this result may be used to distinguish between different models proposed for the superconducting state of $\Sr$.    
\end{abstract}


\maketitle

Chiral superconducting states have attracted an enormous amount of interest in recent years due in large part to their potential for quantum information processing.  They break both parity and time-reversal symmetries and have been predicted to harbor Majorana fermions in vortex cores and along their edges~(see, e.g., \cite{Volovik99,Read00,Stone04}).  The non-Abelian statistics exhibited by these quasiparticles---they are their own antiparticles---endows them with a topological robustness, making them an ideal resource for quantum computation~\cite{Kitaev03}.  To date, one of the most promising candidate chiral superconductors is $\Sr$~\cite{Mackenzie03}.  However, unambiguous evidence of chiral superconductivity is lacking and there is a pressing need to better understand experimental signatures of potential chiral superconductors.  
 The anomalous Hall effect, or the closely related Kerr effect~\cite{Xia06}, is arguably the most direct signature of chiral superconductivity.  However, this effect vanishes in models of clean chiral superconductors studied to date~\cite{Read00,Roy08,Lutchyn08,Goryo08, Lutchyn09}.  

In this Letter, we show that an intrinsic anomalous Hall effect (IAHE) will arise in multiband chiral superconductors provided there is interband pairing
with a \emph{relative phase} (defined below) that differs from that of one (or more) of the intraband order parameters and particle-hole symmetry is broken.  Neither condition is very restrictive and one generally expects any multiband chiral superconductor to satisfy both.   In this case, interband transitions in response to an applied electric field are sensitive to the relative phase of the Cooper pairs, giving rise to a transverse Hall current response.  Using a two-band model of chiral superconductivity, we derive expressions for the frequency dependent Hall conductivity that show this physics explicitly.  

 In general, the orbital part of a 2D chiral order parameter has the form
\beq \Delta_{\alpha}(k_x,k_y) = \Delta^{\prime}_{\alpha}(k_x,k_y)+i\Delta^{\prime\prime}_{\alpha}(k_x,k_y),\label{OP}\eeq
 where $\Delta^{\prime}$ and $\Delta^{\prime\prime}$ are real.  (The global $U(1)$ phase is set to zero).  In a multiband system, there will be multiple order parameters, $\alpha=1,2,...$, possibly arising from both intraband and interband  pairing.  $\Delta_{\alpha}$ is complex and breaks time-reversal symmetry and parity if the real and imaginary parts have different momentum dependencies, such that the Cooper pair electrons have nonzero relative angular momenta.  The momentum-dependent phase $\phi_{\alpha}(\bk) \equiv \tan^{-1}\left(\Delta^{\prime\prime}_{\alpha}(\bk)/\Delta^{\prime}_{\alpha}(\bk)\right)$ plays a central role in characterizing chiral superconductors.   
Responsible for the relative angular momentum between electrons comprising a Cooper pair, we will refer to it as the relative phase of the order parameter throughout.

Although the symmetry of the order parameter is still controversial~\cite{Kallin09}, there is significant experimental evidence that $\Sr$ is chiral $p$-wave~\cite{Mackenzie03}.    One of the strongest pieces of evidence for this is the measurement of a nonzero Kerr angle, $\theta_K \sim 65$ nrads at $T\simeq 0.7K$ ($\simeq 0.45 T_c$)~\cite{Xia06}, an indirect probe of the Hall conductivity $\sigma_H$ at optical frequencies.    The origin and magnitude of this effect is controversial, however, since an ideal (translationally invariant) chiral $p$-wave superconductor would yield $\theta_K=0$~\cite{Read00}. To date, arguably the most promising explanations for the anomalous Hall effect in $\Sr$ have been purely extrinsic, arising from impurity scattering~\cite{Goryo08, Lutchyn09}.   (There is an intrinsic mechanism at finite wavevectors~\cite{Roy08,Lutchyn08}, however this effect would be difficult to probe in experiments.)  Here, we revisit the possibility of an intrinsic contribution in connection with the multiband nature of $\Sr$.   

The Fermi surface of $\Sr$ consists of three cylindrical sheets denoted $\alpha$, $\beta$, and $\gamma$.  The $\gamma$ sheet is an approximately isotropic (in $k_x,k_y$) electron-like Fermi surface while the $\alpha$ and $\beta$ sheets are hole and electron pockets, respectively~\cite{Mackenzie03}.  A number of  analyses of superconductivity in $\Sr$ have concluded that pairing occurs primarily on the $\gamma$ band with passive superconductivity on the $\alpha$ and $\beta$ bands; see, e.g., Refs.~\cite{Agterberg97,Zhitomirsky01,Nomura02,Deguchi04}.   These have assumed only intraband pairing, which, we show, implies (the intrinsic) $\sigma_H=0$.
  
In contrast, a smaller contingent has proposed that superconductivity is strongly multiband, arising primarily on the Ru $d_{xz}$ and $d_{yz}$ orbitals~\cite{Takimoto00,Kuroki01,Annett02,Raghu10}, quasi-1D bands that hybridize to form the $\alpha$ and $\beta$ bands.   This (predominantly) intraorbital pairing gives rise to strong interband pairing and, as a result, a significant IAHE.  Using parameters appropriate for $\Sr$ and a simple $d_{xz}/d_{yz}$ intraorbital pairing model, we find that the intrinsic Hall conductivity yields a Kerr angle on the order of 10-100 nrads at the experimental frequency and low temperatures.   If chiral pairing were to occur primarily on the $\gamma$ band, our analysis suggests that $\sigma_H$ would be strongly suppressed relative to this value.

{\emph{Chiral two-band superconductor---}} Although the choice of single-particle basis used to define the Hamiltonian for a multiband system is irrelevant in the final result for the Hall conductivity, it will be useful to distinguish two bases.  In the ``orbital basis", the Hamiltonian is constructed from microscopic atomic Wannier orbitals, such as the Ru $d$ orbitals in $\Sr$.  In general, there will be an interorbital coupling in this basis and the interorbital contribution to the current is given by the momentum gradient of this coupling.   In contrast, in the ``band basis", the interband current must be found from the interorbital current by unitary transformation.   

We define the Hamiltonian for a two-band system in the orbital basis:
\beq H\! =\! \sum_{\bk}\!\left(\!\begin{array}{cc} c^{\dagger}_{\bk1} & c^{\dagger}_{\bk2}\end{array}\!\right)\left(\!\begin{array}{cc}
\xi_1(\bk)\! & \!\epsilon_{12}(\bk) \\ \epsilon_{12}(\bk) \!& \!\xi_2(\bk)
\end{array}
\right)\left(\begin{array}{c} 
c_{\bk 1} \\ c_{\bk 2}
\end{array}\right) + H_{\mathrm{int}}.\label{H}\eeq
Here, $\xi_{1(2)}\equiv \epsilon_{1(2)}-\mu_{1(2)}$ is the dispersion for the Bloch states constructed from the 1(2) orbital, $\epsilon_{12}$ is the interorbital coupling, and $H_{\mathrm{int}}$ describes interactions, which we assume give rise to intraorbital pairing with order parameters $\Delta_{11}$ and $\Delta_{22}$.  For generality, we include interorbital pairing $\Delta_{12}=\Delta_{21}$, but note that purely intraorbital pairing  ($\Delta_{12}=0$) will still give rise to interband pairing and hence nonzero $\sigma_H$.  

In the  basis defined by the spinor $ \hat{\Psi}^{\dagger}_{\bk} = (c^{\dagger}_{\bk 1},c_{-\bk 1},c^{\dagger}_{\bk 2},c_{-\bk 2})$, the inverse mean-field  $4\times 4$ Green's function for this model is 
\bea \lefteqn{\bG^{-1}_0(\bk,\omega_n) =}&&\nonumber\\&&\!\!\!\!\!\!\!\left(\!\begin{array}{cc}
\!i\omega_n \!-\! \xi_{1}\hat{\tau}_3\!+\! \Delta^{\prime}_{11}\hat{\tau}_1\!-\!\Delta^{\prime\prime}_{11}\hat{\tau}_2 &\! -\epsilon_{12}\hat{\tau}_3 \!+\! \Delta^{\prime}_{12}\hat{\tau}_1\!-\!\Delta^{\prime\prime}_{12}\hat{\tau}_2 \\
\!-\epsilon_{12}\hat{\tau}_3 \!+\! \Delta^{\prime}_{12}\hat{\tau}_1\!-\!\Delta^{\prime\prime}_{12}\hat{\tau}_2&\! i\omega_n \!-\! \xi_{2}\hat{\tau}_3\! +\! \Delta^{\prime}_{22}\hat{\tau}_1\!-\!\Delta^{\prime\prime}_{22}\hat{\tau}_2
\end{array}\!\right).\nonumber\\ \label{G0}\eea
Here, $\hat{\tau}_l$ are the usual $2\times 2$ Pauli matrices,
$\Delta^{\prime}_{ab}$ ($\Delta^{\prime\prime}_{ab}$) is the real (imaginary) part of the intraorbital ($a=b$) and interorbital ($a\neq b$) order parameters, and $\omega_n$ is a Fermi Matsubara frequency.  The two branches of the BCS quasiparticle spectrum, $E_{-}$ and $E_{+}$, are found from the solution of $\mathrm{det}\bG^{-1}_0(\bk,\omega_n) = (\omega^2_n+E^2_{-})(\omega^2_n+E^2_{+})$.

{\emph{Intrinsic Hall conductivity---}  The optical Hall conductivity $\sigma_H(\omega)$ is defined in terms of the antisymmetric part of the $\hat{J}_x$-$\hat{J}_y$ current correlator $\pi_{xy}(\bq,\omega)$ by
\beq \sigma_H(\omega) \equiv -\frac{1}{2i\omega}\lim_{\bq\to 0}\left[\pi_{xy}(\bq,\omega) - \pi_{yx}(\bq,\omega)\right].\label{sigmaH}\eeq
The total current operator in the $i$ direction is given by~\cite{note} $\hat{J}_{i} =e\sum_{\bk}\mathrm{tr}\hat{\Psi}^{\dagger}_{\bk}\hat{\mathbf{v}}_{i}\hat{\Psi}_{\bk}$, where
\beq \hat{\mathbf{v}}_{i} = \left(\begin{array}{cc}
v_{i,11}(\bk)\hat{1}_2  & v_{i,12}(\bk) \hat{1}_2 \\
v_{i,12}(\bk) \hat{1}_2 & v_{i,22}(\bk)\hat{1}_2
\end{array}\right)\label{V}\eeq
is the $4\times 4$ bare current vertex ($\hat{1}_2$ is the $2\times 2$ identity matrix).  In the orbital basis, $v_{i,aa} = \partial_{k_{i}}\epsilon_a$ and  $v_{i,12} = \partial_{k_{i}}\epsilon_{12}$.

Since the intrinsic  Hall effect is essentially a single-particle band effect (although here, the  existence of a time-reversal symmetry-breaking field, $\Delta^{*}_{ab} \neq  \Delta_{ab}$, is a many-body effect), it suffices to evaluate the current correlator at the one-loop level,
\beq \pi_{xy}(\bq,\!\nu_m) \!=\!e^2T\!\!\sum_{\bk,\omega_n}\!\mathrm{tr}[\hat{\mathbf{v}}_x \bG_0(\bk,\!\omega_n)\hat{\mathbf{v}}_y\bG_0(\bk\!+\!\bq,\!\omega_n+\nu_m)],\label{pi0}\eeq
where $\nu_m$ is a Bose Matsubara frequency. In the case of a single orbital (or multiple uncoupled orbitals), $\hat{\mathbf{v}}_{\sigma}$ is purely diagonal and commutes with $\bG_0$.  Consequently, $\pi_{xy}$ equals $\pi_{yx}$, and the one-loop value for the Hall conductivity is zero~\cite{Lutchyn09}.  This result holds independent of details such as band anisotropy and pairing symmetry.  Broken time-reversal symmetry and lack of full translational symmetry are necessary but not sufficient conditions for a nonzero Hall conductivity.  For superconductivity on a single orbital, vertex corrections are crucial to having a nonzero Hall conductivity.  Goryo~\cite{Goryo08} and Lutchyn \textit{et al.}~\cite{Lutchyn09} considered impurity-scattering vertex corrections for a model of superconductivity in $\Sr$ assuming superconductivity takes place predominantly on the $\gamma$ band (or $d_{xy}$ orbital).  

For a multiorbital superconductor, on the other hand, the one-loop contribution (\ref{pi0}) can be nonzero, and if so, should provide a major contribution to the Hall effect in a clean superconductor. This contribution is straightforwardly evaluated by analytically continuing $i\nu_m\to \omega+i0^+$ to real frequencies to obtain the real and imaginary parts of the Hall conductivity (\ref{sigmaH}).  For simplicity, we only show the limiting $T=0$ values:
\bea \lefteqn{\!\!\!\!\!\!\sigma^{\prime}_H(\omega) = 2e^2\sum_{\bk}\frac{\left(\delta\bv_{21}\times\bv_{12}\right)_z}{E_{-}E_{+}(E_{-}+E_{+})((E_{-}+E_{+})^2-\omega^2)}}&&\nonumber\\&&\!\!\!\!\!\!\!\!\!\!\!\!\!\!\times\! \Big[\epsilon_{12}\mathrm{Im}(\Delta^*_{11}\Delta_{22})+\xi_1\mathrm{Im}(\Delta^*_{22}\Delta_{12})-\xi_2\mathrm{Im}(\Delta^*_{11}\Delta_{12})\Big]\label{sigmareal}\eea
and
\bea \lefteqn{\sigma^{\prime\prime}_H(\omega)\! =
\! \frac{\pi e^2}{\omega^2}\sum_{\bk}\frac{\left(\delta\bv_{21}\times\bv_{12}\right)_z}{E_{-}E_{+}}}&&\nonumber\\&&\!\!\!\!\!\!\!\!\!\!\!\!\!\!\times\! \Big[\epsilon_{12}\mathrm{Im}(\Delta^*_{11}\Delta_{22})+\xi_1\mathrm{Im}(\Delta^*_{22}\Delta_{12})-\xi_2\mathrm{Im}(\Delta^*_{11}\Delta_{12})\Big]\nonumber\\&&\times \left[\delta(\omega-E_{-}-E_{+})-\delta(\omega+E_{-}+E_{+})\right].\label{sigmaimag}\eea
Here, $\bv_{ab}\equiv (v_{x,ab},v_{y,ab})$, and $\delta\bv_{21}\equiv \bv_{22}-\bv_{11}$. 

The only terms in the one-loop expression (\ref{pi0}) that contribute to a nonzero $\sigma_H$ are the time-reversal symmetry-breaking interorbital transitions shown in Fig.~\ref{sigmaHfig}.  At $T=0$~\cite{note2}, these transitions involve the creation (or annihilation) of a $\pm$ quasiparticle pair.  The only part of the coherence factors that survive after taking the difference in (\ref{sigmaH}) are terms that involve two different order parameters and connect the two orbitals either through interorbital pairing or hopping.   If the order parameters involved in these processes have different relative phases, i.e., if $\mathrm{Im}(\Delta^{*}_{ab}\Delta_{cd}) = |\Delta_{ab}||\Delta_{cd}|\sin(\phi_{ab}-\phi_{cd})\neq 0$ (see definition below (\ref{OP})), then the interorbital transition will be accompanied by a change in the relative phase of the electrons comprising the quasiparticle pairs.  Because of the asymmetry ($\Delta^{\prime}\neq \Delta^{\prime\prime}$ in (\ref{OP})) between the real and imaginary components of the order parameter with respect to the \emph{relative} momentum $\bk$ of the electrons comprising the Cooper pair, this phase change amounts to a rotation of their relative momentum  in the $k_x,k_y$ plane.  
Since, in a multiband superconductor, the center-of-mass and relative momenta are coupled, this rotation will produce a transverse Hall current.  
Particle-hole asymmetry is required so that the two types of contributions in Fig.~\ref{sigmaHfig}, $(a,b) = (1,2)$ and $(2,1)$, do not cancel.  The sign of $\sigma_H$ also reverses with the chirality  (e.g., $k_x+ik_y \rightarrow k_x-ik_y$).   

\begin{center}
\begin{figure}
\includegraphics[width=0.4\textwidth]{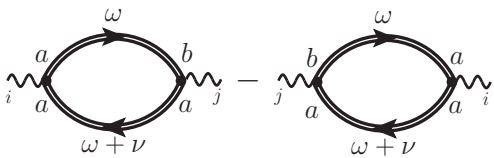}
\caption{Contributions to the intrinsic Hall conductivity in a multiband chiral superconductor in the orbital basis, where $a, b$ label the orbitals and $i,j$ denote the photon polarization.  Double lines denote the  Green's function (\ref{G0}).  At one vertex, a photon of frequency $\nu$ induces a $b\to a$ interorbital transition.  The time-reversed process on the right is subtracted to yield the Hall conductivity.  As discussed in the text, the only nonzero contributions at $T=0$ result from the creation or annihilation of a $\pm$ quasiparticle pair.   At finite $T$, scattering between (but not within) the $\pm$ quasiparticle branches also contribute.
}
\label{sigmaHfig}
\end{figure}
\end{center}

One can transform the velocity vertices and Green's functions entering (\ref{pi0}) into the band basis by sandwiching unitary operators between matrices.  The resulting expression for $\sigma_H$ in this basis has the same form as--and, of course, is equal to--(\ref{sigmareal}) and (\ref{sigmaimag}), but with single-particle energies, velocities and order parameters transformed into the corresponding quantities in the band basis.  Terms in (\ref{sigmareal}) and (\ref{sigmaimag}) proportional to the interband tunneling $\epsilon_{12}$ are absent in the band basis, meaning that $\sigma_H$ is zero unless there is complex interband pairing. 

This conclusion is unchanged when we include spin-orbit coupling (SOC).  For the case where the orbitals are e.g., Ru $d_{xz}$ and $d_{yz}$ orbitals, SOC  is described by 
\beq  H_{\mathrm{SOC}}=i \lambda\sum_{\substack{\bk \eta\eta^{\prime}\\ \sigma\sigma^{\prime}}}c^{\dagger}_{\bk\sigma \eta}c_{\bk\sigma^{\prime}\eta^{\prime}}\epsilon_{\eta\eta^{\prime}l} (\hat{\tau}_l)_{\sigma,\sigma^{\prime}},\label{H0}\eeq
where $\eta=1,2$ enumerates the orbitals, $\sigma$ denotes the two (pseudo)spins, and $\epsilon_{jkl}$ is the totally antisymmetric tensor. Terms proportional to odd powers of $\lambda$ in $\pi_{xy}$  can give rise to a nonzero value of $\sigma_H$ (independent of the chiral order parameters).  However, summing over the two spin species, such terms vanish as long as the Zeeman spin splitting  $\mu_{\uparrow}-\mu_{\downarrow}$ is zero.  
Consequently, SOC only enters $\sigma_H$ indirectly, renormalizing the quasiparticle dispersions.

{\emph{Optical Hall conductivity and Kerr effect in $\Sr$---}} 
$\Sr$ is a three band system with significant SOC which mixes all three $d$ orbitals for wave vectors near $k_x=\pm k_y$~\cite{Haverkort08}.  However, in calculating $\sigma_H$, an important simplification occurs because the current operator only couples $d_{xz}$ and $d_{yz}$ orbitals and hence, only $d_{xz}-d_{yz}$ interorbital transitions contribute to the intrinsic Hall conductivity in $\Sr$.  Consequently, the conclusions reached from our two orbital model are still valid.  In particular, the IAHE in $\Sr$ is only nonzero when there is complex interband pairing and this pairing must  involve the $d_{xz}$ and $d_{yz}$ orbitals to a significant extent.  Because of symmetry, the $d_{xy}$ orbital only plays a passive role in the IAHE, so we simply ignore it and use our two-orbital result for the Hall conductivity, (\ref{sigmareal}) and (\ref{sigmaimag}).
 
Since interorbital coupling plays the key role, we will use the simplest orbital model that neglects next-nearest neighbor intraorbital hopping: $\epsilon_1 =  -2t\cos(k_x a)$, $\epsilon_2  =-2t\cos(k_ya)$, $\epsilon_{12}=2t^{\prime}\sin(k_xa)\sin(k_ya)$.  Recent LDA studies of the band structure of $\Sr$ find $t\simeq 0.4$eV and $t^{\prime} \simeq 0.1t$~\cite{Haverkort08,Rozbicki11}.  
The models of Refs.~\cite{Kuroki01,Annett02,Raghu10} give rise to strong interband pairing on the $d_{xz}$ and $d_{yz}$ orbitals with relative phases differing by $\pi/2$ and comparatively small interorbital and $\gamma$ band pairing~\cite{note3}.  We use $\Delta_{11} \simeq \Delta_0\sin(k_x)$, $\Delta_{22} \simeq i\Delta_0\sin(k_y)$ and $\Delta_{12}=0$, with relative phases $\phi_{22}(\bk)-\phi_{11}(\bk) = \pi/2$.  In this case, (\ref{sigmareal}) can be written as (restoring $\hbar$ and the lattice spacing $a$)
\beq \sigma_H(\tilde{\omega}+i\varepsilon) = (e^2/\hbar)(\tilde{\Delta}_0)^2(\tilde{t}^{\prime})^2 F(\tilde{\mu},\tilde{\Delta}_0,\tilde{\omega}+i\varepsilon)\label{sigmaH2},\eeq
where
\beq F\!=\!-\frac{16}{\pi^2}\!\int_{-\pi}^{\pi}\!\!\!dx dy \frac{\sin^2 x\sin^2 y\left(\cos y\sin^2 x+\cos x\sin^2 y\right)}{ \tilde{E}_1\tilde{E}_2(\tilde{E}_2\!+\!\tilde{E}_1)[(\tilde{\omega}+i\varepsilon)^2-(\tilde{E}_1\!+\!\tilde{E}_2)^2]}\label{F}\eeq
 is dimensionless, with $x\equiv k_x a$, $y\equiv k_ya$, and all quantities with a tilde are scaled by $t$.  To numerically calculate Im$\sigma_H$, we use $\varepsilon = 10^{-6}$.  

 \begin{center}
\begin{figure}
\includegraphics[width=0.42\textwidth]{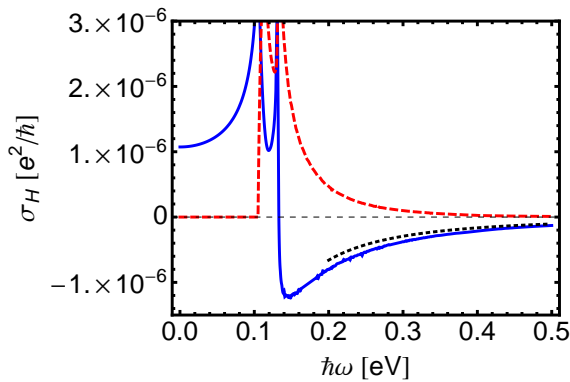}
\caption{(Color online) Real (blue solid line) and imaginary (red dashed line) parts of the $T=0$ Hall conductivity in units of $e^2/\hbar$ as a function of $\hbar\omega$ for $\Delta_0 = 0.23$meV and $t=\mu=10t^{\prime}=0.4$eV. Dotted line shows asymptotic high-frequency expression (see text).}
\label{Sigmaomegafig}
\end{figure}
\end{center}

 \begin{center}
\begin{figure}
\includegraphics[width=0.42\textwidth]{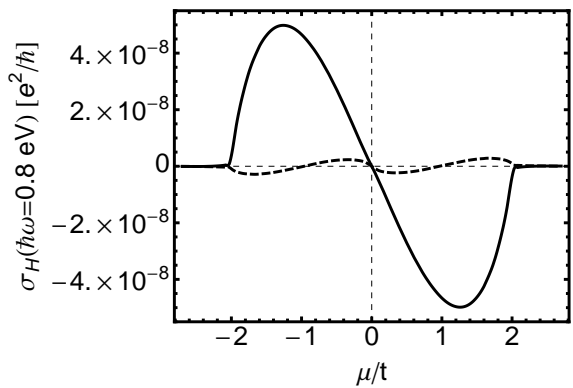}
\caption{Real (solid line) and imaginary (dashed line) parts of the $T=0$ Hall conductivity in units of $e^2/\hbar$ as a function of $\mu/t$ for $\Delta_0 = 0.23$meV and $t=\mu=10t^{\prime}=0.4$eV.}
\label{Sigmamufig}
\end{figure}
\end{center}
  
In Fig.~\ref{Sigmaomegafig}, we plot the real and imaginary parts of $\sigma_H$ as a function of frequency for $t=\mu=0.4$eV and $t^{\prime}=0.1t$.  Following previous studies~\cite{Goryo08,Lutchyn09}, we take $\Delta_0$ to be its BCS value $1.76 T_c$, which is equal to $0.23$meV for the ultraclean samples used in Ref.~\cite{Xia06} with $T_c=1.5$K.  At high frequencies $\hbar\omega \gtrsim 0.4$eV, the (negative) real part dominates the conductivity.  It is well-approximated by the exact asymptotic limit~\cite{Shastry93}, $\sigma_H(\omega\to \infty)= (i/\omega^2)[\langle [\hat{J}_x,\hat{J}_y]\rangle + {\cal{O}}(\omega^{-2})]$, with $\langle [\hat{J}_x,\hat{J}_y]\rangle= -2ie^2T\sum_{\bk}\mathrm{Im}(\sum_{\omega_n}[\bG_0(\bk,\omega_n)]_{13})(\delta\bv_{21}\times\bv_{12})_z$.  Using the above values, this gives $\sigma_H(\omega)\simeq -2.6\times 10^{-8}( e^2/\hbar)/(\hbar\omega/\mathrm{eV})^2$, shown by the dotted line in Fig.~\ref{Sigmaomegafig}. Even though the minimum of the quasiparticle energies lies close to the gap $\Delta_0$, the energy of the quasiparticle pair that determines the imaginary Hall response, $E_{-}(\bk)+E_{+}(-\bk)$, (see (\ref{sigmaimag})) has a minimum around $2t^{\prime}\gg 2 \Delta_0$, accounting for the absence of any spectral weight at  $T=0$ in Im$\sigma_H$ below a value of this order.  

The specific frequency at which the imaginary Hall response becomes nonzero depends on the details of the model, but generically, $t'$ (and/or SOC) separates both the bands and the quasiparticle spectra in energy at fixed wave vector so that the minimum frequency will be of this order and not of order $2\Delta_0$.  If the Fermi surfaces are closer to each other in momentum space (as they may be in a three band model) the structure seen in the imaginary Hall response will shift to somewhat lower frequencies.   We also note that for $T>0$, $+\leftrightarrow -$ quasiparticle transitions fill in some of the low frequency spectral weight.  The rapid rise in Im$\sigma_H$ shown here results from a van Hove singularity for $E_-+E_+$.

In Fig.~\ref{Sigmamufig}, we plot $\sigma_H$ at $\hbar\omega=0.8\mathrm{eV}$, the frequency used in the experiment of Xia \textit{et al.}~\cite{Xia06}, as a function of $\mu/t$.   Fig.~\ref{Sigmamufig} clearly exhibits the need for particle-hole asymmetry ($\mu \neq 0$) discussed earlier.     (For $|\mu|\gtrsim 2t$, our model system is an insulator.)

The Kerr angle $\theta_K(\omega) = (4\pi/\omega d)\mathrm{Im}[\sigma_H/n(n^2-1)]$ depends not only on the Hall conductivity, but also material parameters such as the distance $d$ between Ru-O layers and the complex index of refraction $n(\omega)$~\cite{Lutchyn09}.    Thus, in order to calculate $\theta_K$ one needs knowledge of optical properties of $\Sr$ such as the diagonal component $\sigma(\omega)$ of the conductivity tensor.  Using an experimentally-motivated generalized Drude form for $\sigma(\omega)$ (the same parameters and model as used in Ref.~\cite{Lutchyn09}), we find (see Supplemental Materials) that the intrinsic contribution to the Hall conductivity calculated above gives rise to a Kerr angle of  $\sim 50$ nrads at $\hbar\omega = 0.8$eV.   

Most studies of multiband superconductivity in $\Sr$ assume pairing within the same band, predominantly on the $\gamma$ band with passive pairing on the $\alpha,\beta$ bands~\cite{Agterberg97,Zhitomirsky01,Nomura02,Deguchi04}.  In these models, any interband pairing would likely be substantially suppressed compared to the primary order parameter (on $\gamma$), given the relative sizes of inter- and intraband coupling.  Added to this the fact that the $\gamma$ band only comprises a small admixture of $d_{xz}$ and $d_{yz}$ orbitals~\cite{Haverkort08}, we conclude that the Hall conductivity in these models is likely to be more than an order of magnitude smaller than the estimate we give above.  In contrast, the models of Refs.~\cite{Kuroki01,Annett02,Raghu10}, in which the inter- and intraband order parameters live on the $d_{xz}$, $d_{yz}$ orbitals and have the same magnitude, likely provide the maximum intrinsic Hall conductivity amongst current models of $\Sr$.  

\textit{Conclusions}---In this work, we have shown how an intrinsic, anomalous Hall effect can arise in chiral multiband superconductors provided there is interband pairing and broken particle-hole symmetry---a state of affairs that one would generally expect to be true.  This effect, which has also been studied independently and concurrently in Ref.~\cite{Wysokinski11}, should be generic to all clean multiband chiral superconductors and can provide a powerful optical probe of such systems.  In contrast to previous predictions for intrinsic effects, the interband effect is not restricted to nonzero wavevectors~\cite{Roy08} and does not require the existence of an edge~\cite{Furusaki01} or surface~\cite{Yip92}.  The latter two effects are orders of magnitude smaller than the interband effect. Applying our results to a model of the possible chiral superconductor $\Sr$ in which superconductivity arises primarily on the Ruthenium $d_{xz}$ and $d_{yz}$ orbitals, we find a Hall conductivity of the right order of magnitude to explain Kerr rotation experiments.  As pointed out in Ref.~\cite{Raghu10}, in the absence of superconductivity on the $\gamma$ band, there would \emph{not} be topologically protected Majorana edge modes since the Skyrmion numbers arising from the $\alpha$ and $\beta$ bands  cancel.  However, any chiral superconductivity on the $\gamma$ band, whether induced or arising from microscopic pairing,  would restore the topological nature of the superconducting state at low temperatures.  

By varying the impurity concentration, it is possible that experiments could determine the relative importance of the intrinsic Hall effect described here and extrinsic effects~\cite{Goryo08,Lutchyn09}.  If the intrinsic Hall effect were found to dominate, this would provide compelling evidence for a multiband origin of superconductivity.  

This work was supported by NSERC and the Canadian Institute for Advanced Research.


\newpage

\section{Supplementary Material}

\textit{Calculation of the Kerr angle}---The frequency-dependent  Kerr angle $\theta_K$ is given by (see, for instance, Ref.~\cite{Lutchyn09b})
\beq \theta_K(\omega) = (4\pi/\omega d)\mathrm{Im}[\sigma_H(\omega)\alpha(\omega)],\label{Kerr}\eeq
where $d$ is the interlayer spacing and 
\beq \alpha(\omega) = \frac{1}{n(n^2-1)}.\label{alpha}\eeq
Here, $n(\omega)=\sqrt{\varepsilon_{ab}(\omega)}$ is the complex, frequency-dependent index of refraction, equal to the square root of the component
\beq \varepsilon_{ab} = \varepsilon_{\infty}+(4\pi i/\omega)\sigma(\omega)\label{permeability}. \eeq
of the permeability tensor in the $ab$ plane.  $\varepsilon_{\infty}$ is the background dielectric tensor and $\sigma(\omega)$ is the diagonal element of the conductivity tensor.  

Equations~(\ref{Kerr}) and (\ref{permeability}) show that, in order to extract the frequency-dependent Kerr angle from the optical Hall conductivity $\sigma_H(\omega)$, one also needs to have information about the diagonal element $\sigma(\omega)$ of the optical conductivity tensor.  The only case where any simplification arises is when $\omega$ is much greater than the transport scattering rate $\gamma$.  In this case, the optical conductivity is simply given by the reactive part $\sigma(\omega) \simeq \omega^2_{pl}/4\pi i\omega$, and (\ref{Kerr}) reduces to 
\beq \theta_K(\omega> \omega_{pe}) = \frac{4\pi\omega^2\sigma^{\prime\prime}_H(\omega)}{d\sqrt{\omega^2\varepsilon_{\infty}-\omega^2_{pl}}(\omega^2(\varepsilon_{\infty}-1)-\omega^2_{pl})},\label{Kerr1}\eeq
when $\omega$ is greater than the plasma edge $\omega_{pe}\equiv \omega_{pl}/\sqrt{\varepsilon_{\infty}}$, where $\omega_{pl}$ is the plasma frequency.  Below the plasma edge,  the Kerr angle is dominated by the real part of the Hall conductivity:
\beq \theta_K( \omega<\omega_{pe}) = \frac{4\pi\omega^2\sigma^{\prime}_H(\omega)}{d\sqrt{\omega^2_{pl}-\omega^2\varepsilon_{\infty}}(\omega^2_{pl}-\omega^2(\varepsilon_{\infty}-1))}\label{Kerr2}.\eeq

Although variants of these limiting cases are often used in the literature to estimate the Kerr angle, neither are particularly relevant for the Kerr measurements of Ref.~\cite{Xia06}, which measures the Kerr angle in $\Sr$ at 0.8eV, just below the estimated plasma edge.  Despite the conventional metallic behavior exhibited by $\Sr$ at low temperatures, the optical conductivity of $\Sr$ is not well approximated by a narrow Drude peak~\cite{Katsufuji96}.  The large width of the Drude peak in $\Sr$ leads to a broadening of the response close to the plasma edge and \emph{the Kerr angle receives contributions from both the real and imaginary parts of $\sigma_H$}.  The exact proportion of these contributions depends on the properties of the optical conductivity $\sigma(\omega)$ in this region.  

Kramers-Kronig analyses of reflectivity data~\cite{Katsufuji96} give both the real and imaginary parts of the conductivity $\sigma(\omega)$ at arbitrary frequencies, without any need for modelling.   Such data could thus be used in conjunction with the Hall conductivity $\sigma_H(\omega)$ to directly determine the Kerr angle, or vice-versa.  Absent such information, we 
model the optical conductivity using a Drude model.   While the optical conductivity is better modelled by a generalized Drude expression with a frequency-dependent effective mass and transport scattering rate~\cite{Katsufuji96}, for simplicity, we use a Drude model
\beq \sigma(\omega) =- \frac{\omega^2_{pl}}{4\pi i(\omega + i\gamma)}\label{sigmaDrude}\eeq
with constant transport scattering rate $\gamma$.  The values we use for $\omega_{pe}$ and $\gamma$ are extracted from experiments on $\Sr$.  Following Ref.~\cite{Lutchyn09b}  we take $d=6.8\AA$, $\varepsilon_{\infty}=10$, $\omega_{pl}=2.9\mathrm{eV}$.  From Fig. (4b) in Ref.~\cite{Katsufuji96} (using data for $T=9$K), we estimate $\gamma(0.8\mathrm{eV}) \sim 0.4$, the value we use for all $\omega$.  Using these, in  Fig.~\ref{Kerrfig}, we plot the Kerr angle $\theta_K$ as a function of energy.  At $\omega=0.8$eV, for $\gamma(0.8\mathrm{eV}) = 0.4$eV, we find $\theta_K = 54$nrads.  

 \begin{center}
\begin{figure}
\includegraphics[width=0.37 \textwidth]{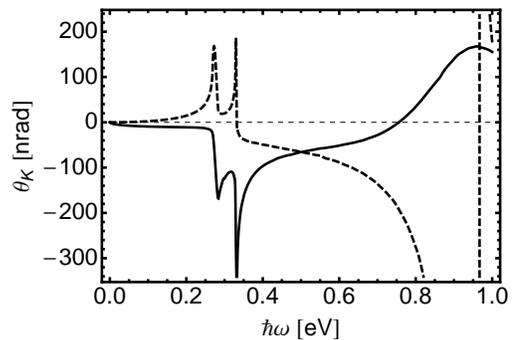}
\caption{Kerr angle as a function of energy for $\Delta_0 = 0.23$meV, $t=0.4$eV, $t^{\prime}/t = 0.1$, $\mu/t = 1$, and $T=0$.  The solid line shows $\theta_K$ for the experimentally-derived transport scattering rate $\gamma = 0.4\mathrm{eV}/\hbar$ estimated from Ref.~\cite{Katsufuji96}.  For comparison, the dashed line shows the often used--but unphysical--approximation $\gamma = 0$.  The experiment of Xia \textit{et al.}~\cite{Xia06} measured the Kerr angle at $\hbar\omega = 0.8$eV.}
\label{Kerrfig}
\end{figure}
\end{center}

We also plot the Kerr angle using (\ref{sigmaDrude}), but assuming $\gamma=0$, i.e., (\ref{Kerr1}) and (\ref{Kerr2}).  This is not a physically realistic limit (except at very high frequencies) and we only show this to emphasize that the simple expressions (\ref{Kerr1}) and (\ref{Kerr2}) should not be used to evaluate the Kerr angle.  The results from using these expressions are evidently very different from those using (\ref{sigmaDrude}) with $\gamma=0.4$eV.


\begin{thebibliography}{99} 
\bibitem{Volovik99} G.~E.~Volovik, JETP Lett. \textbf{70}, 609 (1999). 
\bibitem{Read00} N.~Read and D.~Green, Phys. Rev. B \textbf{61}, 10267 (2000).
\bibitem{Stone04} M.~Stone and R.~Roy, Phys. Rev. B \textbf{69}, 184511 (2004).  
\bibitem{Kitaev03} A.~Y.~Kitaev,  Ann. Phys. \textbf{303}, 2 (2003).
\bibitem{Mackenzie03} A.~P.~Mackenzie and Y.~Maeno, Rev. Mod. Phys. \textbf{75}, 657 (2003).
\bibitem{Xia06} J.~Xia, Y.~Maeno, P.~T.~Beyersdorf, M.~M.~Fejer, and A.~Kapitulnik, Phys. Rev. Lett. \textbf{97}, 167002 (2006).
\bibitem{Lutchyn08} R.~M.~Lutchyn, P.~Nagornykh, and V.~M.~Yakovenko, Phys. Rev. B \textbf{77}, 144516 (2008).  
\bibitem{Roy08} R.~Roy and C.~Kallin, Phys. Rev. B \textbf{77}, 174513 (2008).
\bibitem{Goryo08} J.~Goryo, Phys. Rev. B \textbf{78}, 060501(R) (2008).
\bibitem{Lutchyn09} R.~M.~Lutchyn, P.~Nagornykh, and V.~M.~Yakovenko, Phys. Rev. B \textbf{80}, 104508 (2009).  
\bibitem{Kallin09} C.~Kallin and A.~J.~Berlinsky, J. Phys.: Condens. Matter \textbf{21}, 164210 (2009).
\bibitem{Agterberg97} D.~F.~Agterberg, T.~M.~Rice, and M.~Sigrist, Phys. Rev. Lett. \textbf{78}, 3374 (1997).  
\bibitem{Zhitomirsky01} M.~E.~Zhitomirsky and T.~M.~Rice, Phys. Rev. Lett. \textbf{87}, 057001 (2001). 
\bibitem{Nomura02} T.~Nomura and K.~Yamada,  J. Phys. Soc. Jpn. \textbf{71}, 404 (2002).  
\bibitem{Deguchi04} K.~Deguchi, Z.~Q.~Mao, and Y.~Maeno, J. Phys. Soc. Jpn. \textbf{73}, 1313 (2004). 
\bibitem{Takimoto00} T.~Takimoto, Phys. Rev. B \textbf{62}, R14641 (2000).
\bibitem{Kuroki01} K.~Kuroki, M.~Ogata, R.~Arita, and H.~Aoki, Phys. Rev. B \textbf{63}, 060506(R) (2001).  
\bibitem{Annett02} J.~F.~Annett, G.~Litak, B.~L.~Gy\'{o}rffy, and K.~I.~Wysoki\'{n}ski, Phys. Rev. B \textbf{66}, 134514 (2002).  
\bibitem{Raghu10} S.~Raghu, A.~Kapitulnik, and S.~A.~Kivelson, Phys. Rev. Lett. \textbf{105}, 136401 (2010).
\bibitem{note} Although we suppress spin in our model, the Hall conductivity receives equal contributions from both spins.  The current operator includes contributions from both particle and hole contributions, mimicking spin counting.  Hence, the Hall conductivity calculated from this spin operator correctly accounts for two spins.  
\bibitem{note2} At nonzero $T$, there are also processes involving quasiparticles-quasiparticle transitions between the two branches. 
\bibitem{Haverkort08} M.~W.~Haverkort, I.~S.~Elfimov, L.~H.~Tjeng, G.~A.~Sawatzky, and A.~Damascelli, Phys. Rev. Lett. \textbf{101}, 026406 (2008).
\bibitem{Rozbicki11} E.~J.~Rozbicki, J.~F.~Annett, J.-R.~Souquet, and A.~P.~Mackenzie, J. Phys.: Condens, Matter \textbf{23}, 094201 (2011).  Note that our factor of $t^{\prime}$ is equivalent to twice $f_{xz/yz}$ in this reference.  
\bibitem{note3} Ref.~\cite{Takimoto00} concludes that only one of $d_{xz}$, $d_{yz}$ exhibits strong pairing, which in our analysis would lead to a smaller Hall conductivity. 
\bibitem{Shastry93} B.~S.~Shastry, B.~I.~Shraiman, and R.~R.~P.~Singh, Phys. Rev. Lett. \textbf{70}, 2004 (1993).
\bibitem{Wysokinski11} K.~I.~Wysokinski, J.~F.~Annett, and B.~L.~Gyorffy, arXiv:1111.5309. 
\bibitem{Furusaki01} A.~Furusaki, M.~Matsumoto, and M.~Sigrist, Phys. Rev. B \textbf{64}, 054514 (2001).
\bibitem{Yip92} S.~K.~Yip and J.~A.~Sauls, J. Low Temp. Phys. \textbf{86}, 257 (1992). 
\end{thebibliography}

\begin{thebibliography}{99} 
\bibitem{Lutchyn09b} R.~M.~Lutchyn, P.~Nagornykh, and V.~M.~Yakovenko, Phys. Rev. B \textbf{80}, 104508 (2009).  
\bibitem{Xia06} J.~Xia, Y.~Maeno, P.~T.~Beyersdorf, M.~M.~Fejer, and A.~Kapitulnik, Phys. Rev. Lett. \textbf{97}, 167002 (2006).
\bibitem{Katsufuji96} T~Katsufuji, M.~Kasai, and Y.~Tokura, Phys. Rev. Lett. \textbf{76}, 126 (1996).  
\end{thebibliography}
\end{document}